\documentclass[showpacs,amssymb,floatfix]{revtex4}
\usepackage{graphicx}
\usepackage{dcolumn}
\usepackage{bm}

\begin{document}
\date{\today}
\title{Electric circuit networks equivalent to chaotic quantum billiards}
\author{Evgeny N. Bulgakov$^{1,2}$, Dmitrii N. Maksimov$^1$ and Almas F. Sadreev$^{1,2,3}$}
\affiliation{1) Kirensky Institute of Physics, 660036,
Krasnoyarsk, Russia\\ 2) Astaf'ev Pedagogical University, 660049,
Lebedeva, 89, Krasnoyarsk, Russia\\ 3) Department of Physics and
Measurement, Technology Link\"{o}ping  University,  S-581 83
Link\"{o}ping, Sweden}

\begin{abstract}
We formulate two types of electric RLC resonance network equivalent
to quantum billiards. In the network of inductors grounded by
capacitors squared resonant frequencies are eigenvalues of the
quantum billiard. In the network of capacitors grounded by
inductors squared resonant frequencies are given by inverse eigen
values of the billiard. In both cases local voltages play role of
the wave function of the quantum billiard. However as different
from quantum billiards there is a heat power because of resistance
of the inductors. In the equivalent chaotic billiards we derive
the distribution of the heat power which well describes numerical
statistics.
\end{abstract}
\maketitle
\section{Introduction}
Firstly electric circuit models representing a quantum particle in
the one-dimensional potential
\begin{equation}\label{schreq}
 -\frac
{\hbar^2}{2m}\frac{\partial^2\psi(x)}{\partial
x^2}+V(x)\psi(x)=E\psi(x)
\end{equation}
 were considered by Kron in 1945
\cite{kron}. Three types of equivalent circuits were established.
The first one contains positive and negative resistors and in each
state the currents and voltages are constant in time. The second
and third models are similar and consist of inductors and
capacitors and the currents and voltages are sinusoidal in time.
Here we consider the stationary Schr\"odinger equation in
two-dimensional billiards in hard wall approximation
\begin{equation}\label{laplace}
  -\nabla^2 \psi(x,y)=\epsilon \psi(x,y),
\end{equation}
where the Dirchlet boundary condition is implied at the boundary
$C$ of the billiard:
\begin{equation}\label{DB}
  \psi_{|{_C}}=0.
\end{equation}
Here we use Cartesian coordinates $x, y$ which are dimensionless
via a characteristic size of the billiard $L$, and correspondingly
$\epsilon=\frac{E}{E_0}, E_0=\frac{\hbar^2}{2mL^2}$.

 There is a complete equivalence of the two-dimensional
Schr\"odinger equation for a particle in a hard wall box to
microwave billiards \cite{stockmann}. A wave function is exactly
corresponds to the the electric field component of the TM mode of
electromagnetic field: $\psi(x,y)\leftrightarrow E_z(x,y)$ with
the same Dirichlet boundary conditions. This equivalence is turned
out very fruitful and allowed to test a mass of predictions found
in the quantum mechanics of billiards \cite{stockmann}. On the
other hand, models for the equivalent RLC circuit of a resonant
microwave cavity exist which establish the analogy near an
eigenfrequency \cite{sucher}. Manolache and Sandu \cite{manolache}
proposed a model of resonant cavity associated to an equivalent
circuit consisting of an infinite set of coupled RLC oscillators.
Therefore, there to be a bridge between quantum billiards and the
set of coupled RLC oscillators \cite{berggren1}. In fact, we show
here that at least, two models of electric resonance circuits
(ERC) can be proposed. In the first model shown in Fig. \ref{fig1}
the eigen wave functions correspond correspond to voltages and
eigen energies do to squared eigen frequencies of ERC. In the
second model shown in Fig. \ref{fig2} the eigen energies of
quantum billiard correspond to the inverse squared eigen
frequencies of the electric network. The electric network analogue
systems allow to measure not only typically quantum variables such
as probability and probability current distributions but also a
distribution of heat power in chaotic billiards. Moreover
intrinsic resistances of the RLC circuit allow to model the
processes of decoherence.
\section{Electric resonance circuits equivalent to quantum
billiards}
   If to map the two-dimensional Schr\"odinger equation onto numerical grid
$(x,y)=a_0(i,j), i=1,2,\ldots N_x, j=1,2,\ldots N_y$ one can
easily obtain equation in finite element approximation
\begin{equation}\label{finelem}
\psi_{i,j+1}+\psi_{i,j-1}+\psi_{i+1,j}+\psi_{i-1,j}+(a_0^2E-4)\psi_{i,j}=0.
\end{equation}
The equivalent Hamiltonian is the tight-binding one
\begin{equation}\label{tight}
  H=-\sum_{i,j}\sum_{{\bf b}}|ij\rangle\langle|ij+{\bf b}|,
\end{equation}
where vector ${\bf b}, |{\bf b}|=1$ runs over the nearest
neighbors.

\begin{figure}[ht]
\includegraphics[width=0.5\textwidth]{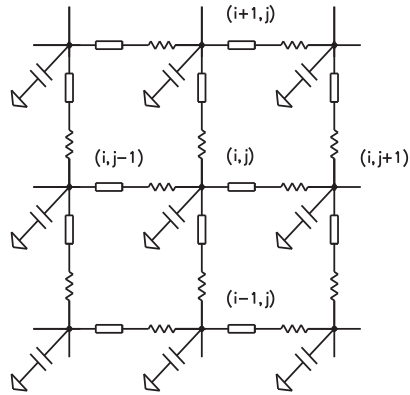}
\caption{The first model of resonance RLC circuits.} \label{fig1}
\end{figure}

Let us consider the electric resonance circuit shown in Fig.
\ref{fig1}. Each link of the two-dimensional network is given by
the inductor L with the impedance
\begin{equation}\label{zL}
z_L=i\omega L+R
\end{equation}
where $R$ is the resistance of the inductor and $\omega$ is the
frequency. Each site of the network is grounded via the capacitor
$C$ with the impedance
\begin{equation}\label{zC}
  z_C=\frac{1}{i\omega C}.
\end{equation}
\begin{figure}[ht]
\includegraphics[width=.55\textwidth]{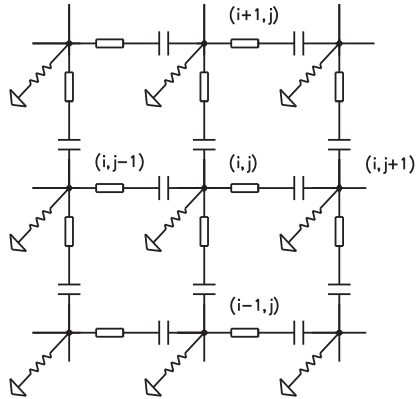}
\caption{The second model of resonance RLC circuits.} \label{fig2}
\end{figure}

The Kirchoff's current law at each site of the network gives
\begin{equation}\label{kirchoff}
  \frac{1}{z_L}[V_{i,j+1}-V_{i,j}+V_{i,j-1}-V_{i,j}+V_{i+1,j}-V_{i,j}
  +V_{i-1,j}-V_{i,j}]-\frac{1}{z_C}V_{i,j}=0,
\end{equation}
where $V_{i,j}$ are values of voltage at the site $(i,j)$. One can
see that this equation coincides with the discretized version of
the Schr\"odinger equation (\ref{finelem}) with the eigenenergies
as
\begin{equation}\label{eigenenergy}
  a_0^2k^2=-\frac{z_L}{z_C}=LC\omega^2-iRC\omega=
  \frac{\omega^2}{\omega_0^2}-i\frac{\gamma\omega}{\omega_0^2},
\end{equation}
where $\omega_0=1/\sqrt{LC}$ and $\gamma=R/L$ are the eigen
frequency and the linewidth of each resonance circuit.

For the second network of electric resonance circuits shown in
Fig. \ref{fig2} we obtain
\begin{equation}\label{kirchoff2}
  \frac{1}{z_C}[V_{i,j+1}-V_{i,j}+V_{i,j-1}-V_{i,j}+V_{i+1,j}-V_{i,j}
  +V_{i-1,j}-V_{i,j}]-\frac{1}{z_L}V_{i,j}=0.
\end{equation}
Therefore, comparing with (\ref{finelem}) we have
\begin{equation}\label{eigenenergy2}
  a_0^2k^2=-\frac{z_C}{z_L}=\frac{1}{LC\omega^2}+\frac{iR}{L\omega}=
  \frac{\omega_0^2}{\omega^2}+i\frac{\gamma\omega_0^2}{\omega}.
\end{equation}
where  $\gamma=RC$.
This network is interesting in that
its eigen frequencies are inverse to the eigenenergies of the
quantum billiard.

There are many ways to define the boundary conditions (BC). Let it
be $(i_B, j_B)$ are sites which belong to the boundary of the
network. If these sites are grounded, we obtain obviously the
Dirichlet BC (\ref{DB}) $V_{\vert_B}=0$. It they are shunted
through capacitors we obtain the free BC (the von Neumann BC). At
last, if the boundary sites are shunted through resistive
inductors, the BC correspond to mixed BC.
\section{Analog of the chaotic Bunimovich billiard}

\begin{figure}[ht]
\includegraphics[width=.6\textwidth]{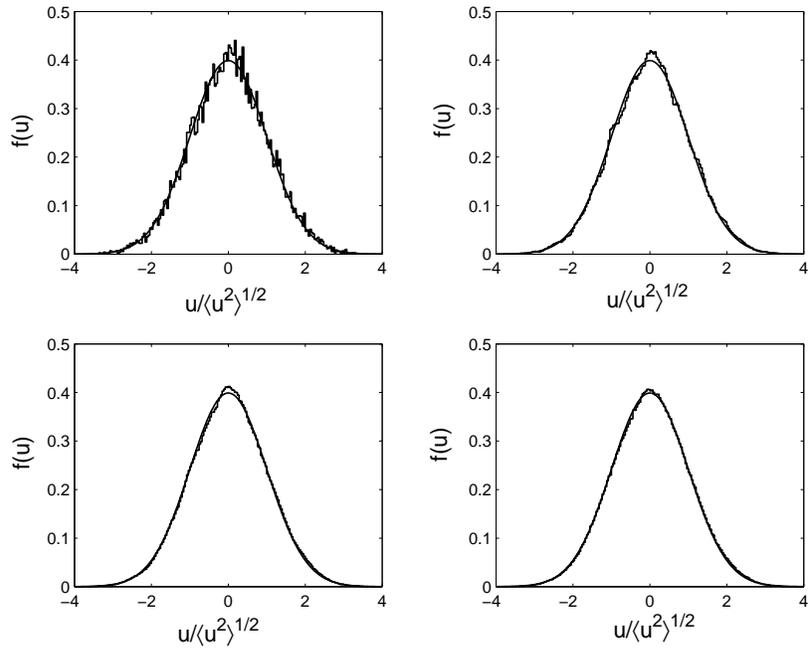}
\caption{The distribution of real part of the wave function of the
quarter Bunimovich billiard mapped on resonance RLC circuit with
elementary unit $a_0=0.01, \omega=1.722 MHz , L=0.1~ mH , C=1~ nF , R=0$.
(a) There is no tolerance of the
electric circuit elements. (b) The tolerance equals to $1\%$.
(c) The tolerance equals to $3\%$.
(d) The tolerance equals to $5\%$. Each distribution in (b) - (d)
is averaged over 100 realizations of the electric network. }
\label{fig3}
\end{figure}

A real electric circuit network has three features which can make
a difference if to compare to the quantum billiards. These are 1)
a discreteness of  resonance circuits, 2) tolerance of electric
elements,  and 3) resistance of inductors.  In practice the
discreteness has no effect for $\lambda \geq 10 a_0$ where
$\lambda$ is a characteristic wavelength of  wave function, and
$a_0$ is the elementary unit of the network. Numerically we
consider the electric network with shape as a quarter of the
Bunimovich billiard.  The distribution of real part of the wave
function of the billiard mapped on the electric circuit network
with $a_0=1/100$ is shown In Fig. \ref{fig3} (a). The wavelength
$\lambda=2\pi a_0\omega_0/\omega=0.115$ for parameters given in
caption of Fig. \ref{fig3}.
 We take the width of the billiard as
unit. One can see distinctive deviation from the Gaussian
distribution which is result of multiple interference on discrete
elements of the network.

It is known that a noise, for example, temperature, smoothes the
conduction fluctuations for transmission through quantum billiards
\cite{iida,prigodin}. In present case the tolerance of circuit elements,
capacity and inductance, plays role of the noise. Therefore we can
expect that increasing of the tolerance can suppress fluctuations
of the distribution of the wave function of the discrete electric
circuit network. In fact, even the $1\%$ tolerance substantially
smoothes the distribution of the wave function as shown in Fig.
\ref{fig3} (b)-(d). We consider that the fluctuations of
capacitors and inductors are not correlated at different sites.

\begin{figure}[ht]
\includegraphics[width=.34\textwidth]{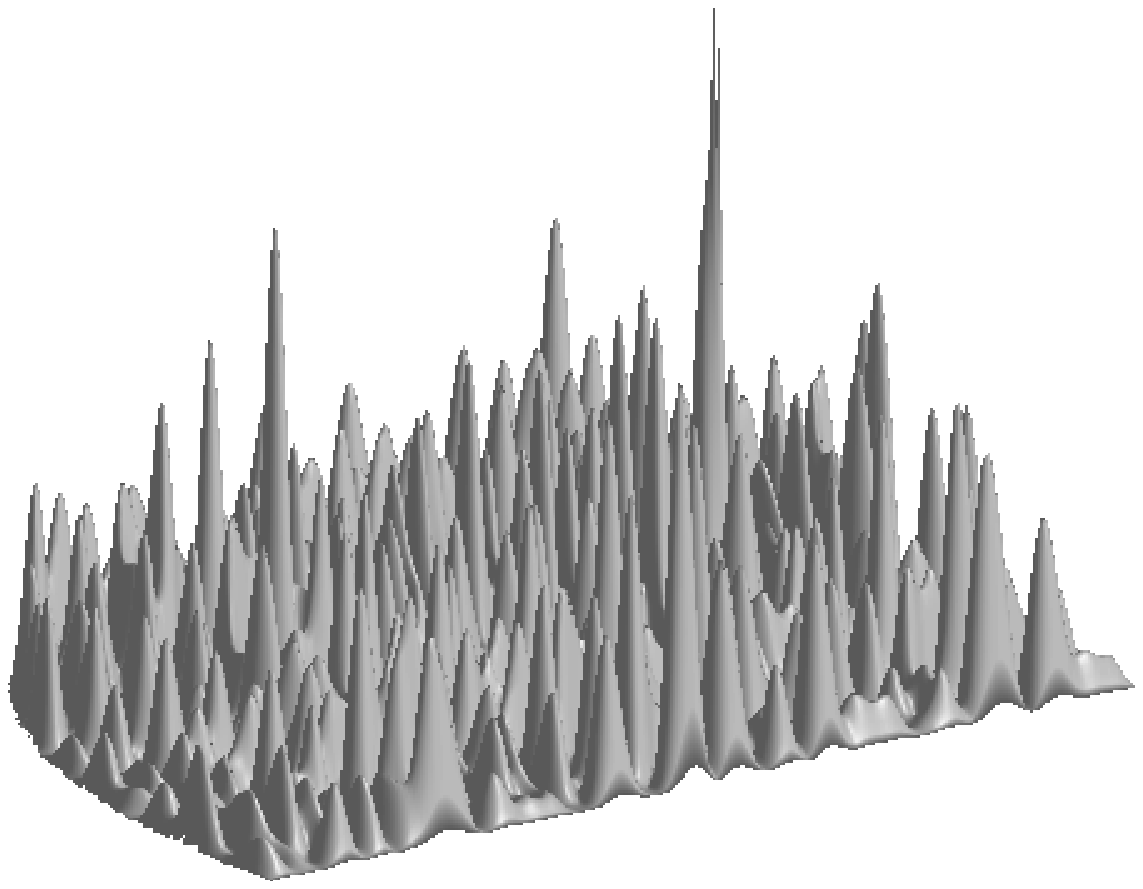}
\includegraphics[width=.34\textwidth]{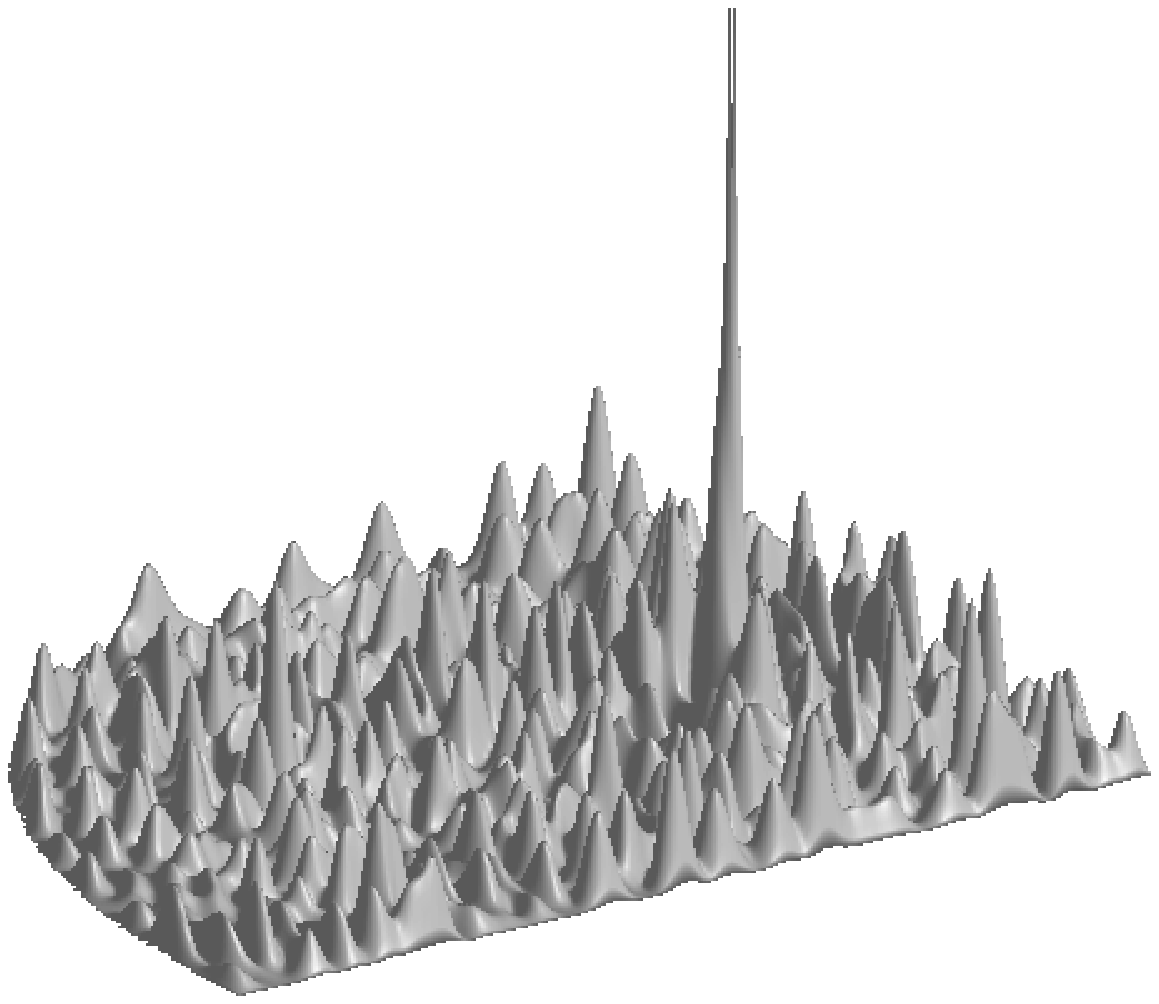}
\caption{Views of probability density  of the quarter
Bunimovich billiard
mapped on resonance RLC circuit with elementary unit $a_0=0.005$,
 $\omega=0.8611 ~MHz, L=0.1~ mH , C=1~ nF$. Left $R=0.5~\Omega$, right $R=1~\Omega$.
 The point of connection
 of external ac current is at maximum of the probability density.} \label{fig4}
\end{figure}

Finally we consider as a damping caused by resistance of the
inductors effects the distribution of the wave function in the
electric RLC resonance circuit network. In order to excite the
network we apply external ac current at single site of the
network. Fig. \ref{fig4} shows the probability density in the
quarter of the Bunimovich billiard for two values of the
resistance $R$. One can see from Fig. \ref{fig4} (right) a
localization effect because of a damping of the probability
density flowing from ac source (see also Fig. \ref{streamline}).
The characteristic length of space damping can be easily estimated
from Eq. (\ref{eigenenergy}) which gives us
\begin{equation}
\lambda_R \approx \frac{4\pi a_0}{R}\sqrt{\frac{L}{C}}.
\end{equation}

\begin{figure}[ht]
\includegraphics[width=.6\textwidth]{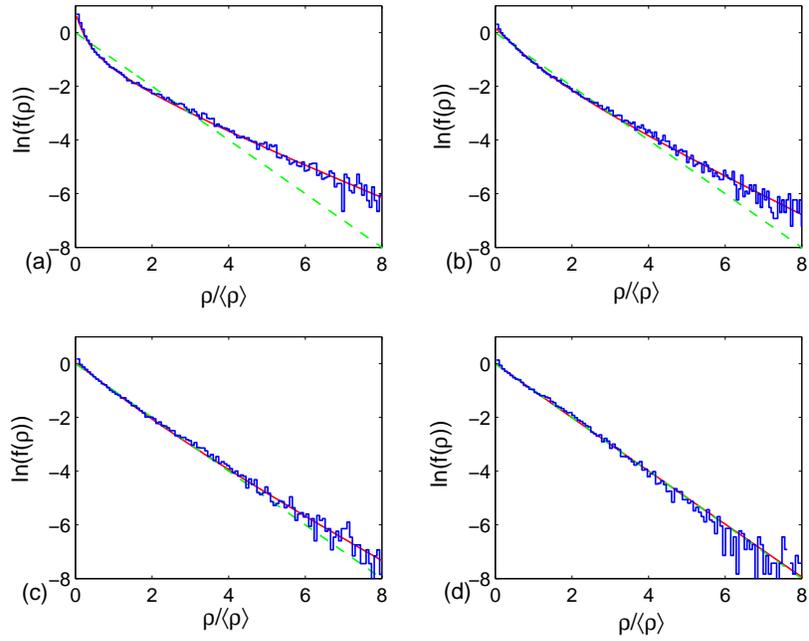}
\caption{(Color online) Distribution of probability density  of
the quarter Bunimovich billiard mapped on resonance RLC circuit
with the same parameters as given in Fig. \ref{fig4}. (a)
$R=0.1~\Omega, Q=3162, \epsilon=0.2488$, (b) $R=0.3~\Omega,
Q=1054, \epsilon=0.5308$, (c) $R=0.5~\Omega, Q=632,
\epsilon=0.6996$ and (d) $R=1~\Omega, Q=316, \epsilon=0.9164$. The
distribution (\ref{probability}) is shown by solid red line, the
Rayleigh distribution $f(\rho)=\exp(-\rho)$ is shown by dashed
green line. } \label{fig5}
\end{figure}

The distributions of the probability density $\rho=|V|^2$ for open
quantum chaotic billiards were considered in many articles
\cite{lenz,lenz1,kanzieper,pnini,ishio} for the case of zero
damping. Here we follow \cite{saichev,ishio} and perform the phase
transformation $V\rightarrow V\exp(i\theta)=p+iq$ which makes the
real and imaginary parts of the wave function $V$ independent.
Introducing a parameter of openness of the billiard \cite{saichev}
\begin{equation}\label{epsilon}
  \epsilon^2=\frac{\sigma_q^2}{\sigma_p^2}.
\end{equation}
where $\sigma_p^2=\langle p^2\rangle, ~ \sigma_q^2=\langle
q^2\rangle$ we can write the distribution of probability density
as \cite{ishio}
\begin{equation}\label{probability}
  f(\rho)=\mu\exp(-\mu^2\rho)I_0(\mu\nu\rho),
\end{equation}
with the following notations
\begin{equation}\label{munu}
  \mu=\frac{1}{2}\left(\frac{1}{\epsilon}+\epsilon\right),
\nu=\frac{1}{2}\left(\frac{1}{\epsilon}-\epsilon\right),
\end{equation}
and $I_0(x)$ is the modified Bessel function of zeroth order, This
distribution is shown in Fig. \ref{fig5} by solid lines while the
Rayleigh distribution $f(\rho)=\exp(-\rho)$ is shown by dashed
lines. The Rayleigh distribution specifies the distribution of
completely open system. One can see from Fig. \ref{fig5} (a, b)
that the statistics of the probability density follows the
distribution (\ref{probability}) irrespective to resistance $R$.
However with growth of the resistance the distribution
(\ref{probability}) tends to the Rayleigh distribution (Fig.
\ref{fig5} (c, d)). Since the larger resistance the more quantum
system is open, this tendency of statistics of the probability
density is clear.

\begin{figure}[ht]
\includegraphics[width=.6\textwidth]{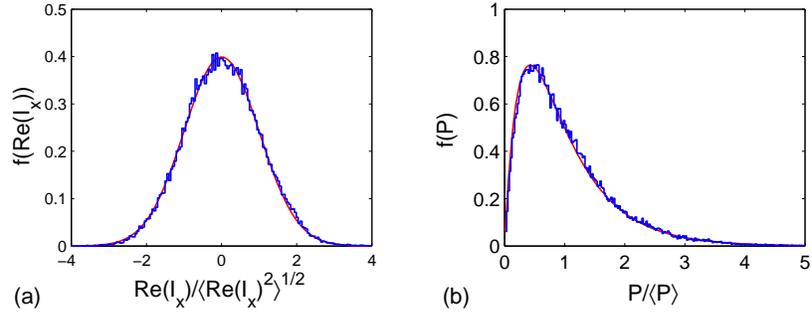}
\caption{(Color online) (a) Statistics of the real part of the
x-component of electric current $I_x$ compared to the Gaussian
distribution shown by solid red line. (b) Statistics  of the heat
power compared to the distribution \ref{fPeps} shown by solid red
line. Here the quarter Bunimovich billiard is taken with
$\omega=1.163~ MHz, R=0.1~ \Omega, L=0.1~ mH , C=1~ nF$.}
\label{fig6}
\end{figure}
\section{The heat power}

In open systems the probability current density corresponds to the
Poynting vector. The last equivalence allowed to test in
particular universal current statistics in chaotic billiards
\cite{barth,saichev}. However in the electric resonance circuit
there is the heat losses because of the resistance. A local power of the
heat losses is defined by formula \cite{heat}
\begin{equation}\label{heat}
  P=\frac{R}{2}[Re(I_x)^2+Im(I_x)^2+Re(I_y)^2+Im(I_y)^2]=
  \frac{R}{2}[|I_x|^2+|I_y|^2],
\end{equation}
where $I_x, I_y$ are local components of the electric power
flowing between sites of the electric network:
\begin{equation}\label{heat power}
  RI_x(i,j)=V_{i+1,j}-V_{i,j},\quad RI_y(i,j)=V_{i,j+1}-V_{i,j}.
\end{equation}
If to approximate the true state with the Berry conjecture
\begin{equation}\label{berry}
  V(x,y)=\sum_j a_j\exp[i~({\bf k_j r}+\phi_j)]
\end{equation}
where $a_j$ and $\phi_j$ are independent random real variables and
$k_j$ are randomly oriented wave vectors of equal length, then $V$
is the complex random Gaussian field (RGF) in the chaotic
Bunimovich billiard. The derivatives of $V$ are also independent
complex RGFs. The components $I_x, ~I_y$ form two complex RGFs
with the probability density of these fields
\begin{equation}
\label{fIxIy}
 f(I_x',I_y',I_x'',I_y'')=
\frac{1}{4\pi^2\sigma_r^2\sigma_i^2}\exp\left\{-\frac{1}{2}
\left(\frac{I_x'^2+I_y'^2}{\sigma_r^2}+\frac{I_x''^2+I_y''^2}{\sigma_i^2}\right)\right\}
\end{equation}
where $I_x'=Re(I_x), ~I_y'=Re(I_y), ~I_x''=Im(I_x),
~I_y''=Re(I_y), ~\sigma_r^2=\langle I_x'^2\rangle,\langle
I_y'^2\rangle, ~\sigma_i^2=\langle I_x''^2\rangle,\langle
I_y''^2\rangle$. In numerical computations we use that average over
the billiard area
\begin{equation}\label{averdef}
  \langle \ldots \rangle = \frac{1}{A}\int d^2{\bf x} \ldots,
\end{equation}
is equivalent to average over three complex GRFs
\begin{equation}\label{aver}
  \langle \ldots\rangle=\int d^2V d^2I_x d^2I_y f(Re(V),Im(V))
f(I_x',I_y',I_x'',I_y'')\ldots.
\end{equation}
An example of the distribution of the real part of $I_x$ is
presented in Fig. \ref{fig6} (a) which shows that numerically this
value is, in fact,  the RGF. A definition of the probability
distribution (\ref{fIxIy}) is relied on that the Berry function
(\ref{berry}) is isotropic in space : $\langle
I_x'^2\rangle=\langle I_y'^2\rangle, ~\langle
I_x''^2\rangle=\langle I_y''^2\rangle$. In fact, an anisotropy of
the shape of billiard effects an anisotropy. However this effect
is the boundary condition's one which has of order
$L_P\lambda/A\sim \lambda $, where $L_P$ is a length of the
billiard perimeter, and $\lambda$ is a characteristic wave length
of wave function in terms of the width of the billiard. Therefore,
for the excitation of the eigenfunction with sufficiently high
frequency we can use the distribution function (\ref{fIxIy}).
Table 1 of numerically computed  mean values confirms this
conclusion.\\ \ \\ Table 1. Numerically computed mean values. \\ \
\\
\begin{tabular}{|c|c|c|c|c|c|} \hline
 $\omega$, MHz & the wavelength $\lambda $
 in terms of the billiard's width& $\frac{\langle I_x'^2\rangle-\langle
I_y'^2\rangle}{\langle I_x'^2\rangle+\langle
I_y'^2\rangle}$ & $\frac{\langle I_x''^2\rangle-\langle
I_y''^2\rangle}{\langle I_x''^2\rangle+\langle
I_y''^2\rangle}$ & $\epsilon$ \cr \hline
0.8611 & 0.1154 & 0.095  & -0.128 &0.2488 \cr \hline
1.1623 & 0.0854 &  0.056 &0.050 & $0.6103$  \cr \hline
\end{tabular}\\ \ \\

To find the distribution of the heat power (\ref{heat}) it is
convenient to begin with a characteristic function
\begin{equation}\label{char0}
  \Theta(a)=\langle\exp(iaP)\rangle=\int d^2I_x d^2I_y
f(I_x',I_y',I_x'',I_y'')\exp(iaR[|I_x|^2+|I_y|^2]/2).
\end{equation}
Substituting (\ref{fIxIy}) we obtain
\begin{equation}\label{char}
\Theta(a)=-\frac{(\sigma_r^2+\sigma_i^2)^2}{\sigma_r^2\sigma_i^2}
\frac{1}{\left(a+i\frac{\sigma_r^2+\sigma_i^2}{\sigma_r^2}\right)
\left(a+i\frac{\sigma_r^2+\sigma_i^2}{\sigma_i^2}\right)}.
\end{equation}
A knowledge of the characteristic function allows to find the heat
power distribution function
\begin{equation}\label{fP}
  f(P)=\frac{1}{2\pi}\int_{-\infty}^{\infty}da\Theta(a)\exp(-iaP)=
  \frac{2\mu }{\nu\langle P\rangle}\exp(-\mu P/\langle P\rangle)\sinh(\nu P/\langle
  P\rangle),
\end{equation}
where formulas (\ref{munu} take the following form
\begin{equation}\label{munu1}
  \mu=\frac{(\sigma_r^2+\sigma_i^2)^2}{2\sigma_r^2\sigma_i^2},\quad
\nu=\frac{(\sigma_r^2-\sigma_i^2)^2}{2\sigma_r^2\sigma_i^2}.
\end{equation}
For $\sigma_r^2\approx \sigma_i^2$ the distribution takes the very
simple form
\begin{equation}\label{fPsimple}
  f(P)=\frac{4P}{\langle P\rangle^2}\exp(-2P/\langle P\rangle),
\end{equation}
Even for this case the distribution of heat power differs from the
distribution of the probability current \cite{saichev}. The
parameter of openness of the billiard (\ref{epsilon}) can be
approximated as
\begin{equation}\label{epsilon1}
  \epsilon^2=\frac{\sigma_i^2}{\sigma_r^2}.
\end{equation}
It is easy to obtain from the Schr\"odinger equation that
$2\sigma_r^2=E\sigma_p^2, ~2\sigma_i^2=E\sigma_q^2$ from which the
last equality in (\ref{epsilon1}) follows.

 Then the heat power distribution function
(\ref{fP}) can be written as follows
\begin{equation}\label{fPeps}
  f(P)=\frac{1+\epsilon^2}{1-\epsilon^2}\left\{\exp\left(-\frac{(1+\epsilon^2)P}
{\langle P\rangle}\right)-\exp\left(-\frac{(1+\epsilon^2)P}
{\epsilon^2\langle P\rangle}\right)\right\}.
\end{equation}
This distribution is shown in Fig. \ref{fig6} (b) which as one can
see nicely describes numerically computed statistics of the heat
power.
If to introduce a value
\begin{equation}\label{sigmaP}
  \sigma_P^2=\frac{\langle (P-\langle P\rangle)^2\rangle}{\langle
  P\rangle^2},
\end{equation}
then one can derive the relation between this parameter and the
parameter of openness (\ref{epsilon1})
\begin{equation}\label{relation}
  \sigma_P^2=\frac{\epsilon^4+1}{(\epsilon^2+1)^2}.
\end{equation}
If the quantum system is fully opened, $\epsilon=1$, and we have
from (\ref{relation}) that $\sigma_P^2=1/2$. For the limit of
closed quatum system we obtain correspondingly that
$\sigma_P^2=1$.
\section{summary and conclusions}
\begin{figure}[ht]
\includegraphics[width=.7\textwidth]{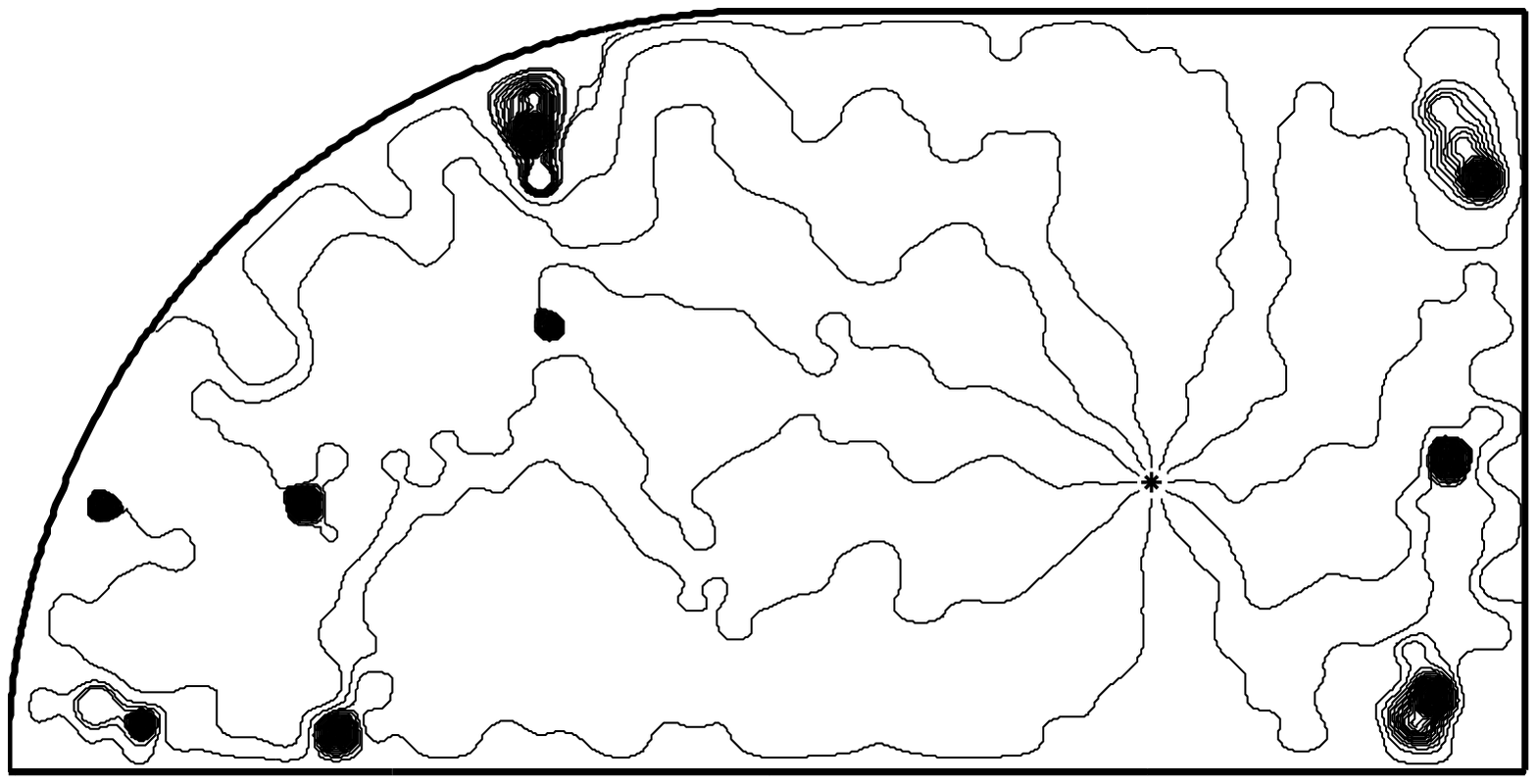}\\ \ \\
\includegraphics[width=.5\textwidth]{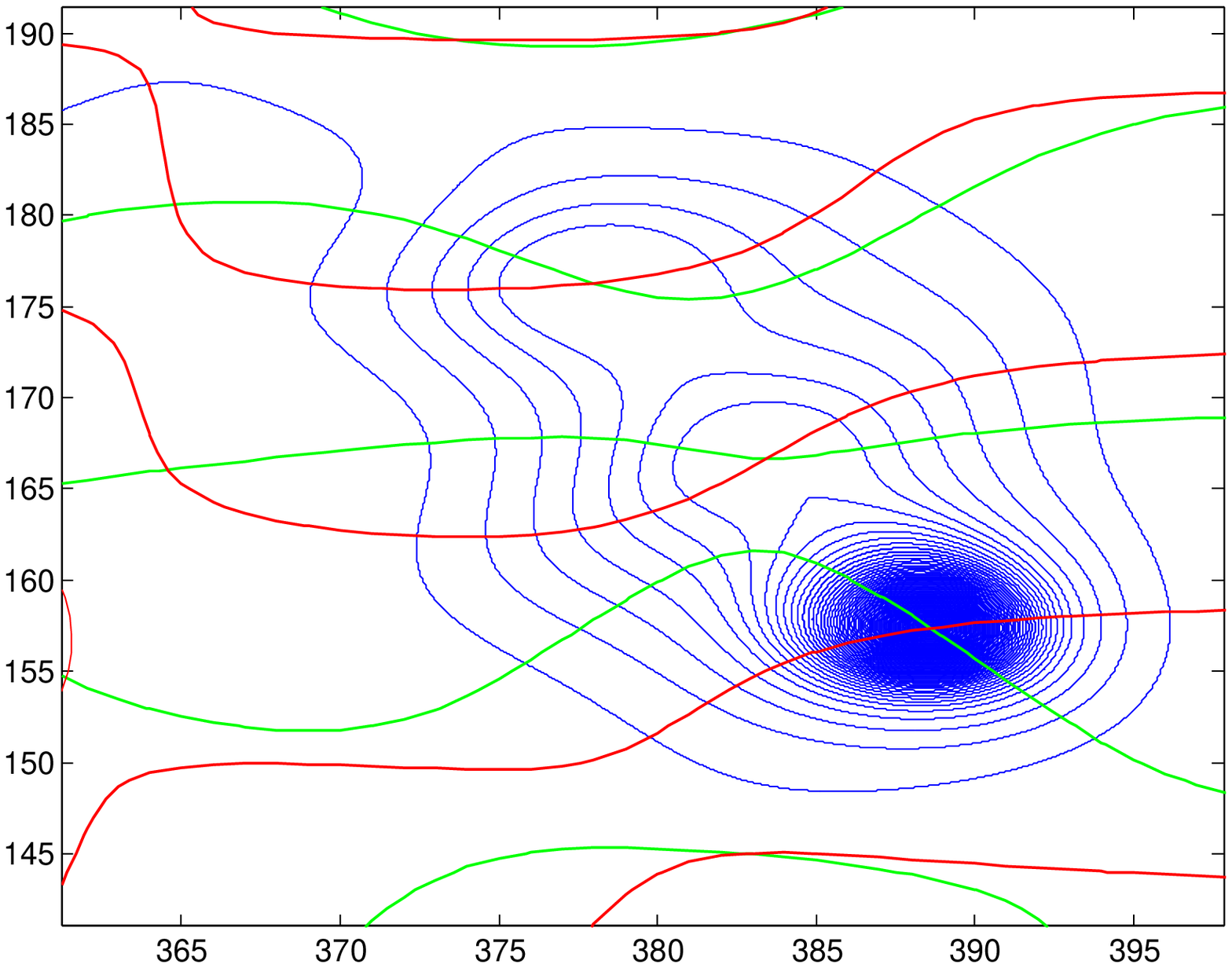}
\caption{Top: quantum streamlines in the quarter of the Bunimovich
billiard flowing from point shown by star at which the external ac
current is applied. (Color online) Bottom: zoomed part of top
figure. Blue lines show the streamline, red and green lines are
the nodal lines of the real and imaginary parts of wave function
correspondingly. The points at which the nodal lines intersecting
are centers of the vortices \cite{berggren}. The wave function
corresponds to Fig. \ref{fig4} (right) with the same parameters.}
\label{streamline}
\end{figure}

We established two types of the electric circuit networks of the
RLC resonant oscillators in which voltages play role of quantum
wave function. Specifically we considered the electric networks
with the Dirichlet boundary conditions which are equivalent to the
quarter of the Bunimovich quantum billiard. In fact, the electric
circuit network has three features which can make a difference if
to compare with the quantum billiards. These are a discreteness of
resonance circuits, tolerance of electric elements,  and
resistance of inductors. We showed numerically that the first two
features are conceal each other. The resistance of the electric
network gives rise to a heat which can be described locally by the
heat currents. Assuming that the wave function in the billiard can
be given as the complex random Gaussian field we derived the
distribution of the heat power which well describe numerical
statistics.

The third feature of the electric network, resistance has
principal importance. The resistance of the electric network is
originated from inelastic interactions of electrons with phonons
and other electrons which give rise to irreversible processes of
decoherence.  With growth of the resistance the wave function is
becoming localized. We studied as the probability density and the
probability currents evolves with increasing of the resistance.
Therefore we can conclude that the resistance violates the
equation $\nabla {\bf j}=0$. In fact Fig. \ref{streamline}
 demonstrates unusual behavior of quantum
streamlines \cite{berggren} with growth of the resistance. One can
see that the quantum streamline are terminating at the vortex
cores. The vortices serve as sinks for the probability density
shown in  Fig. \ref{streamline} (top) as spots. Therefore the
resistance of the inductors in the equivalent electric networks is
simple mechanism of a deterioration of ballistic transport similar
to the B\"uttiker mechanism \cite{buttiker}.

\acknowledgments AS is grateful to K.-F. Berggren for numerous
fruitful discussions.  This work by Russian Foundation for Basic
Research (RFBR Grants 05-02-97713, 05-02-17248). AS acknowledges support of the
Swedish Royal Academy of sciences.

\end{document}